\documentclass[conference]{IEEEtran} 
\IEEEoverridecommandlockouts

\usepackage[british]{babel}
\usepackage[utf8x]{inputenc}

\usepackage{times}

\usepackage{graphicx}

\usepackage[numbers,sort&compress]{natbib}

\usepackage{pgf}
\usepackage{tikz}
\usetikzlibrary{positioning}
\tikzstyle{line} = [draw, -latex]
\tikzstyle{noarrow} = [draw, -]
\usepackage[RPvoltages]{circuitikz}

\usepackage{amsmath, amsfonts, amsthm, amssymb}

\usepackage{times}
\usepackage{graphicx}
\usepackage{subfigure}
\usepackage{url}
\usepackage{xcolor}

\usepackage{balance}

\makeatletter
\def\blfootnote{\xdef\@thefnmark{}\@footnotetext}
\makeatother

\DeclareGraphicsExtensions{.png,.eps,.ps,.pdf}

\usepackage[nolist, nohyperlinks]{acronym}

\begin{acronym}

\acro{PLS}{physical layer security}
\acro{ASC}{average secrecy capacity}
\acro{RV}{random variable}
\acro{SNR}{signal-to-noise ratio}
\acrodefplural{SNR}[SNRs]{signal-to-noise ratios}
\acro{AWGN}{additive white Gaussian noise}
\acro{CSI}{channel state information}
\acro{PDF}{probability density function}
\acrodefplural{PDF}[PDFs]{probability density functions}
\acro{CDF}{cumulative distribution function}
\acro{cCDF}{complementary CDF}
\acrodefplural{CDF}[CDFs]{cumulative distribution functions}
\acro{BS}{base station}
\acro{TDD}{time-division duplexing}
\acro{UL}{uplink}
\acro{DL}{downlink}
\acro{MRT}{maximal ratio transmission}
\acro{MRC}{maximal ratio combining}
\acro{MC}{Monte Carlo}
\acro{MIMO}{multiple-input multiple-output}
\acro{CTA}{Technological Corporation of Andalusia}
\acro{TAS}{transmit antenna selection}
\acro{O-TAS}{optimal transmit antenna selection}
\acro{B-TAS}{Bob-based transmit antenna selection} 
\acro{E-TAS}{Eve-based transmit antenna selection} 
\acro{IoT}{Internet of Things}
\acro{RF}{radio-frequency}
\acro{BS}{base station}

\end{acronym}

\begin{document}


\title{A New Transmit Antenna Selection Technique for Physical Layer Security with Strong Eavesdropping\\
\thanks{978-1-6654-0306-1/21/\$31.00 \copyright2021 IEEE.}
}



\author{\IEEEauthorblockN{Gonzalo J. Anaya-L\'opez}
\IEEEauthorblockA{\textit{Instituto de Telecomunicaci\'on (TELMA)} \\
\textit{Universidad de Málaga}\\
Málaga 29010, Spain \\
gjal@ic.uma.es}
\and
\IEEEauthorblockN{J.~Carlos~Ruiz-Sicilia}
\IEEEauthorblockA{\textit{Dept. of  Communication Engineering} \\
\textit{Universidad de Málaga}\\
Málaga 29071, Spain \\
jcrs@ic.uma.es}
\and
\IEEEauthorblockN{F. Javier~L\'opez-Mart\'inez}
\IEEEauthorblockA{\textit{Instituto de Telecomunicaci\'on (TELMA)} \\
\textit{Universidad de Málaga}\\
Málaga 29010, Spain \\
fjlopezm@ic.uma.es}
}

\maketitle

\begin{abstract}
We propose a new \ac*{TAS} technique that can be beneficial for physical layer security purposes. Specifically, we show that the conventional TAS criterion based on the legitimate \ac*{CSI} is not recommended when the average signal-to-noise ratio for the illegitimate user becomes comparable or superior to that of the legitimate user. We illustrate that an eavesdropper's based antenna selection technique outperforms conventional TAS, without explicit knowledge of the eavesdropper's instantaneous CSI. Analytical expressions and simulation results to support this comparison are given, showing how this new TAS scheme is a better choice in scenarios with a strong eavesdropper.
\end{abstract}


\section{Introduction} \label{sec:introduction}

Security in wireless communications is a critical issue due to the inherent broadcast nature of the wireless channel. Cryptographic techniques based on complex mathematical functions must be updated as computing power advances \cite{LiuRuoheng2010SWCa}. In addition, the advent of the \ac{IoT} with energy and power limited devices demands strategies that do not incur in a high computational burden or complex key distribution protocols with high network overheads. For this reason, the information theoretic approaches based on \ac{PLS} have become an alternative to provide security in wireless environments \cite{Shannon1949,Bloch2008}.

\Ac{PLS} techniques make use of the physical characteristics of the wireless channel in order to achieve the desired security. Wyner in \cite{wyner1975wire} proved the existence of codes for the wiretap channel that guarantee security with this complementary technique to the traditional approach. Later, in \cite{Leung1978} this work was extended to the Gaussian wiretap and the secrecy capacity was defined as the difference between the main channel capacity and the eavesdropper's one, being the former capacity always greater than the latter.

Motivated by the advances of multiple antennas systems in wireless communication, complex beamforming techniques were considered \cite{mukherjee2010robust, sheng2018beamforming} to improve channel security. However, this solution requires a higher complexity and additional computational cost that increases with the number of antennas \cite{sanayei2004antenna, Mohammadi2012}. Hence, we propose the alternative \ac{TAS} \cite{sanayei2004antenna}, which requires only one \ac{RF} chain and therefore reduces the hardware complexity \cite{alves2011enhanced, alves2012performance}.

\Ac{TAS} techniques are useful to reduce the hardware complexity of transmitters while retaining some of the benefits conveyed by multi-antenna transmission \cite{sanayei2004antenna}. For this reason, they have also been exploited in the context of \ac{PLS}. Classically, \ac{TAS} strategies for \ac{PLS} rely on a sub-optimal selection criterion that is solely based on the \ac{CSI} of the legitimate link \cite{alves2012performance,yang2012transmit,wang2014b,Moualeu}. The optimal \ac{TAS} criterion that maximizes the achievable secrecy rates requires perfect \ac{CSI} knowledge for both the legitimate and eavesdropper's links \cite{sadeque2013average}, which may not be always possible in practice, and incurs in an additional complexity penalty.

In this work, we investigate whether conventional \ac{TAS} techniques are always the best choice from a \ac{PLS} perspective. Specifically, we evaluate the performance of an eavesdropper's based \ac{TAS} scheme. This technique can be implemented with low feedback and low signal processing cost since, as in the legitimate's based \ac{TAS} scheme \cite{yang2012transmit}, only the index of the worst antenna for the Eve's channel is required. We show that this new criterion outperforms conventional \ac{TAS} techniques as those in \cite{alves2012performance,yang2012transmit,wang2014b,Moualeu} in the \textit{strong eavesdropper} regime \cite{vincentpoorstrongeve}; i.e., when the average \ac{SNR} for the eavesdropper is comparable or larger than the legitimate \ac{SNR}.

\begin{figure}[t]
	\centering
      \begin{tikzpicture}[node distance = 7cm, auto, place/.style={circle,draw=black,fill=black,thick,radius=2pt}, scale=1, every node/.style={transform shape}]

        \node [draw, minimum width=2cm, minimum height=1cm, text centered,label] (A) {Alice};
        \node [draw, minimum width=2cm, minimum height=1cm, right of=A] (B) {Bob};
        \node [draw, minimum width=2cm, minimum height=1cm, below=1.5cm of B] (E) {Eve};

        \filldraw[black] ([xshift=0.25cm, yshift=0.125cm]A.east) circle (0.5pt);
        \filldraw[black] ([xshift=0.25cm]A.east) circle (0.5pt) node [right] {\scriptsize M};
        \filldraw[black] ([xshift=0.25cm, yshift=-0.125cm]A.east) circle (0.5pt);
        \path [noarrow]([yshift=0.25cm]A.east) -- ++ (0.5,0) node[antenna,scale=0.5] {};
        \path [noarrow]([yshift=-0.25cm]A.east) -- ++ (0.8,0) node[antenna,scale=0.5] {};
        \path [line]([xshift=1cm]A.east) -> node [above] {${y_B =  h^{(k)}_{\rm B} z_{\rm B} + n_{\rm B}}$} (B);
        \path [line]([yshift=-0.25cm, xshift=1cm]A.east) -> node [yshift=-0.1cm] {${y_E = h^{(k)}_{\rm E} z_{\rm B} + n_{\rm E}}$} (E.west);
      \end{tikzpicture}
	\caption{System model under consideration. The DL transmission of the message $z_{\rm B}$ by the BS (Alice) takes place through the transmit antenna $k$, which is selected according to some \ac{TAS} criterion.}
	\label{fig:model}
\end{figure}
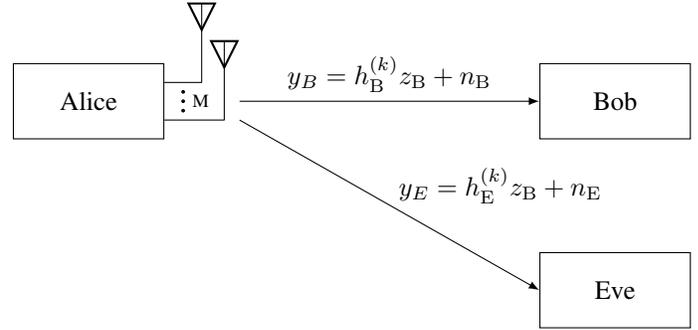

\section{System Model} \label{sec:systemmodel}
We consider a mobile communications scenario as in Fig. \ref{fig:model}, where the \ac{BS} is equipped with $M$ antennas while the receivers are single-antenna users. The \ac{BS} is equipped with a single \ac{RF} chain, and hence uses \ac{TAS} to exploit the spatial diversity due to the use of a multi-antenna configuration. We also take the conventional assumption that all wireless channels are affected by independent quasi-static fading, which are constant during the transmission of an entire codeword. In order to communicate with the set of served users $\mathcal{V}$ within its coverage area, a \ac{TDD} protocol is considered. Due to channel reciprocity, \ac{CSI} from each user can be acquired from the \ac{UL} transmissions.

For the \ac{DL} transmission, the \ac{BS} can operate in two modes, which we denote as \emph{secure} and \emph{standard} modes. In the former, the \ac{BS} aims to communicate securely with a certain user $v_{\rm i}=\text{B}$, for which the placeholder name Bob is conventionally used. At the same time, a different user (eavesdropper, or Eve) $v_{\rm j}=\text{E}$ aims to intercept and decode the communication, taking advantage of the broadcast nature of the wireless transmission. In the latter mode, the \ac{BS} transmits a set of messages $z_{v}$ with $\mathbb{E}\{|z_v|^2\}=1$ to each and every user $v\in\mathcal{V}$. Regardless of the \ac{DL} transmission mode, the \ac{BS} uses a \ac{TAS} scheme with a certain antenna selection criterion.


Now, focusing on the \ac{DL} transmission in secure mode, the \ac{BS} transmits the message $z_{\rm B}$ with  $\mathbb{E}\{|z_{\rm B}|^2\}=1$ through the $k$-th selected antenna according to the \ac{TAS} scheme. Therefore, the signal received at Bob 
from the BS in the \ac{DL} transmission is given by
\begin{align}
    y^{\text{TAS}}_{\rm B} &=  \sqrt{P_T R_{\rm B}^{-\alpha}} {{h}_{\rm B}^{\rm TAS}} z_{\rm B}+n_{\rm B},
    \label{eq:yB}
\end{align}
whereas the signal received at the eavesdropper is given by
\begin{align}
    y^{\text{TAS}}_{\rm E} &=  \sqrt{P_T R_{\rm E}^{-\alpha}} {{h}_{\rm E}^{\rm TAS}} z_{\rm B}+n_{\rm E},
    \label{eq:yE}
\end{align}
where $P_T$ denotes the transmit power, 
$n_{\rm B}$ and $n_{\rm E}$ are the \ac{AWGN} components at each receiver, with $\mathbb{E}\{|n_{{\rm B},{\rm E}}|^2\}=N_0$, $R_{\rm B}$ and $R_{\rm E}$ are the distances between the \ac{BS} and Bob/Eve, respectively, and $\alpha$ is the path-loss exponent. The channel coefficients ${{h}_{u}^{\rm TAS}}=h^{(k)}_{u}$, with $u=\{{\rm B,E}\}$, correspond to those between the $k$-th transmit antenna and each receiver, so that ${{h}_{u}^{\rm TAS}}\in\left[h^{(1)}_{u},...,h^{(M)}_{u}\right]$ and the channel coefficients $h_{\rm u}^{(i)}$ are normalized as $\mathbb{E}\{|h_{\rm u}^{(i)}|^2\}=1$.


From \eqref{eq:yB} and \eqref{eq:yE} we can define the instantaneous \acp{SNR} at the legitimate and eavesdropper's sides as
\begin{align}
    \gamma_{\rm B} &= \frac{P_T R_{\rm B}^{-\alpha}}{N_0}\left|h_{\rm B}^{\rm TAS}\right|^2\left|z_{\rm B}\right|^2
    \label{eq:snrB}
\end{align}
and
\begin{align}
    \gamma_{\rm E} &=  \frac{P_T R_{\rm E}^{-\alpha}}{N_0}\left|h_{\rm E}^{\rm TAS}\right|^2\left|z_{\rm B}\right|^2.
    \label{eq:snrE}
\end{align}

As aforementioned, we first assume that perfect \ac{CSI} is available for every user in the system, which is acquired during the \ac{UL} transmission; this condition will be later relaxed. Hence, the maximum reliable and secure transmission rate is defined by the \ac{ASC}:
\begin{align}
    \overline{C}_{\rm S} = \mathbb{E}\{C_{\rm S}\left(\gamma_{\rm B}, \gamma_{\rm E}\right)\},
    \label{eq:avCs}
\end{align}
where $C_{\rm S}\left( \gamma_{\rm B}, \gamma_{\rm E}\right)$ is the instantaneous secrecy capacity defined as
\begin{align}
    {C}_{\rm S}( \gamma_{\rm B}, \gamma_{\rm E})\underset{\gamma_{\rm B}> \gamma_{\rm E}}{=}\left[\log_2(1+ \gamma_{\rm B})-\log_2(1+ \gamma_{\rm E})\right]^+,
    \label{eq:iCs}
\end{align}
where $\left[\cdot \right]^+$ is used to denote that ${C}_{\rm S}=0$ when the argument within brackets is negative. When a random antenna is selected for transmission, the coefficients $h_{\rm B}^{\rm TAS}$ and $h_{\rm E}^{\rm TAS}$ have the same distribution as if a single-antenna \ac{BS} was considered, i.e., there is no gain having multiple-antennas at the \ac{BS}. However, this situation changes when a \ac{TAS} criterion other than random is considered. In the following subsections, we discuss how the use of different \ac{TAS} schemes affects the achievable secrecy rates defined in \eqref{eq:iCs}.

\subsection{Optimal TAS}\label{sec:otas}
The \ac{O-TAS} scheme selects the transmit antenna that provides the best instantaneous transmission rate for the system \cite{sadeque2013average}, i.e., maximizes the instantaneous secrecy capacity in \eqref{eq:iCs}. To this end, it is required perfect \ac{CSI} of the legitimate and the eavesdropper channel. The average \acp{SNR} for these channels is given as follows:
\begin{align}
    \overline\gamma_{\rm B}
    =\frac{P_T R_{\rm B}^{-\alpha}}{N_0}\mathbb{E}\left\{\left|h_{\rm B}^{\rm TAS}\right|^2\right\}=\overline\gamma_{\rm B_0}\mathbb{E}\left\{\left|h_{\rm B}^{\rm TAS}\right|^2\right\}
    \label{eq:avsnrb}
\end{align}
and
\begin{align}
    \overline\gamma_{\rm E}=
    \frac{P_T R_{\rm E}^{-\alpha}}{N_0}\mathbb{E}\left\{\left|h_{\rm E}^{\rm TAS}\right|^2\right\}=\overline\gamma_{\rm E_0}\mathbb{E}\left\{\left|h_{\rm E}^{\rm TAS}\right|^2\right\},
    \label{eq:avsnre}
\end{align}
where $\overline\gamma_{\rm B_0}$ and $\overline\gamma_{\rm B_0}$ 
are the average \acp{SNR} in the case of a single-antenna transmitter. With these definitions, we can obtain the $k$-th antenna that maximizes the rate of secure transmission using the following selection criterion:
%
\begin{align}
    k=\underset{1\leq i\leq M}{\arg \max}\Bigg\{\log_2\Bigg(\underbrace{\tfrac{1+\overline\gamma_{\rm B_0} \left|h^{(i)}_{\rm B}\right|^2}{1+\overline\gamma_{\rm E_0} \left|h^{(i)}_{\rm E}\right|^2}}_{X_i}\Bigg)>0\Bigg\}.
    \label{eq:criterion}
\end{align}

Hence, the selection criterion aims at maximizing the random variable $X_i$, built as a generalized ratio of Bob's and Eve's instantaneous \acp{SNR}. This is only possible when perfect \ac{CSI} for all channels is available at the \ac{BS}, yielding:

\begin{align}
    \overline C_{\rm S}^{\rm O-TAS}=\mathbb{E} \Bigg\{\log_2\Bigg({\tfrac{1+\overline\gamma_{\rm B_0} \left|h^{(k)}_{\rm B}\right|^2}{1+\overline\gamma_{\rm E_0} \left|h^{(k)}_{\rm E}\right|^2}}\Bigg)>0\Bigg\}.
   \label{eq:otas}
\end{align}


\subsection{Bob-based TAS}\label{sec:btas}
The \ac{B-TAS} scheme is the state-of-the-art technique usually associated with \ac{TAS} \cite{alves2012performance,yang2012transmit,wang2014b,Moualeu} in the \ac{PLS} literature. Under this criterion, the antenna that maximizes the legitimate channel transmission rate is selected, instead of maximizing the secrecy capacity. For this reason, it is regarded as a sub-optimal scheme \cite{zhu2015secrecy}. Despite not selecting the best possible antenna from a secrecy perspective, this scheme is popular in practice compared to \ac{O-TAS} because it does not require eavesdropper's \ac{CSI}. Besides, full Bob's \ac{CSI} is also not required to implement this \ac{TAS} scheme, since the antenna index information can be retrieved from Bob using a low-rate feedback channel, or estimated at Alice using an energy detector. With this scheme, the antenna selection criterion is simpler, as it is just a function of the legitimate channel coefficients:
\begin{align}
    h_{\rm B}^{\rm B-TAS}=\max\limits_{i=1,...,M} \left\{ \left|h^{(i)}_{\rm B}\right|^2 \right\},
    \label{eq:btas}
\end{align}
where the distribution of $h_{\rm B}^{\rm B-TAS}$ is defined by the maximum (or $M$-th order) statistic \cite{david2004order} as follows when $h^{(i)}_{\rm B}\, \forall i=1,...,M$ are independent and identically distributed:
\begin{align}
    F_{\rm B}^{M}(x) = F(x)^M,
    \label{eq:btascdf}
\end{align}
where $F(x)$ is the \ac{CDF} of a single fading coefficient. Note that the distribution in \eqref{eq:btascdf} depends on the number of antennas $M$ among which the selection is made. Conversely, the distribution of $h_{\rm E}^{\rm B-TAS}$ is not altered by the \ac{B-TAS} selection criterion.



\subsection{E-TAS}\label{sec:etas}
The last scheme to analyze, which is originally presented in this work to the best of the authors' knowledge, is referred to as \ac{E-TAS}. This scheme aims to improve the system's secrecy performance by selecting the transmit antenna solely based on the eavesdropper's channel. The rationale behind this scheme comes from the observation that \ac{B-TAS} criterion tends to behave as the \ac{O-TAS} criterion as the eavesdropper's \ac{SNR} is reduced, i.e. $\overline\gamma_{\rm E_0} \left|h^{(i)}_{\rm E}\right|^2\ll 1$ in \eqref{eq:otas}, since the latter becomes immaterial in the maximization process. Based on this observation, we conjecture that an \ac{E-TAS} criterion aimed at minimizing the rate of Eve's channel should perform reasonably close to \ac{O-TAS} when the eavesdropper's channel has a better average \ac{SNR} than the legitimate channel; this is the case, for instance, of scenarios with near users \cite{zhang2020physical} or strong eavesdroppers \cite{vincentpoorstrongeve}. In other words, it is the denominator in \eqref{eq:otas} that would dominate in the antenna selection. 

Secrecy capacity can be improved by either increasing the capacity of the legitimate channel or decreasing the capacity of the eavesdropper channel. Therefore, the \ac{E-TAS} scheme improves secrecy capacity by selecting the antenna that presents the \emph{worst} channel for the eavesdropper. In this way, the legitimate's \ac{CSI} is not required in the selection criterion:
\begin{align}
    h_{\rm E}^{\rm E-TAS}=\min\limits_{i=1,...,M} \left\{ \left|h^{(i)}_{\rm E}\right|^2 \right\},
    \label{eq:etas}
\end{align}
where $h_{\rm E}^{\rm E-TAS}$ now follows the distribution defined by the minimum (or first-order) statistic \cite{david2004order} as follows when $h^{(i)}_{\rm E}\, \forall i=1,...,M$ are independent and identically distributed:
\begin{align}
    F_{\rm E}^{1}(x) = 1-\left[1-F(x)\right]^{M},
    \label{eq:etascdf}
\end{align}
while the distribution of $h_{\rm B}^{\rm E-TAS}$ is not changed by the \ac{E-TAS} selection criterion.
Note that in our scheme the eavesdropper's \ac{CSI} is available at the \ac{BS} from the \ac{UL} transmission phase; however, this is not explicitly required to implement the antenna selection criterion. Instead, a simple antenna-wise energy detection technique at the \ac{BS} during the \ac{UL} transmission is enough to implement the \ac{E-TAS} criterion. Although targeting the eavesdropper is unusual, our aim is to demonstrate how in scenarios where perfect \ac{CSI} is not available, choosing the \ac{B-TAS} criterion can be far from an optimal decision.

\section{Physical Layer Security}\label{sec:pls}

We now evaluate the average secrecy capacity for the \ac{E-TAS} scheme, as the figure of merit to compare the secrecy performances. Evidently, in all instances the performance upper-bound will be provided by the \ac{O-TAS} scheme, since the antenna selection criterion incorporates both Bob's and Eve's \ac{CSI}. In the following, for the sake of simplicity and ease of discussion, we assume the case of Rayleigh fading.

For the \ac{O-TAS} scheme, we gently refer to the results obtained in \cite[eq. (13)]{sadeque2013average}, which have a rather complicated form due to the intricate nature of the distribution of the equivalent \ac{RV} $X_i$ in \eqref{eq:criterion}.

In order to obtain the expression of the sub-optimal schemes in a simpler way, we use the \ac{ASC} formulation in \cite[eq. (12)]{Wang2014}:
\begin{align}
    \overline{C}_{\rm S}\left(\overline\gamma_{\rm B},\overline\gamma_{\rm E}\right)=\frac{1}{\ln 2}\int_0^{\infty}\frac{F_{\rm E}(x)\left[1-F_{\rm B}(x)\right]}{1+x}dx,
    \label{eq:Cs}
\end{align}
where $F_{\rm B}(\cdot)$ and $F_{\rm E}(\cdot)$ represent the \acp{CDF} of $\gamma_{\rm B}$ and $\gamma_{\rm E}$, respectively.

Since we assumed complex Gaussian distributed channel coefficients, we have that the channel coefficients not affected by the selection criteria, i.e., $h_{\rm E}^{\rm B-TAS}$ and $h_{\rm B}^{\rm E-TAS}$, are exponentially distributed with \ac{CDF}:
\begin{align}
    F_{\rm exp}(x,\beta) = 1-\exp\left(\frac{-x}{\beta}\right),
    \label{eq:eCDF}
\end{align}
where $\beta$ is the mean of an exponential distribution.

Substituting \eqref{eq:btascdf} and \eqref{eq:eCDF} in \eqref{eq:Cs}, applying the binomial theorem and after some manipulations, we obtain the \ac{ASC} for the \ac{B-TAS} scheme: 
\newcommand\A{\frac{k}{\overline\gamma_{\rm B_0}}}
\newcommand\B{{\frac{1}{\overline\gamma_{\rm E_0}} + \A}}
\begin{align}
    \overline{C}^{\rm B-TAS}_{\rm S} = &\frac{1}{\ln 2}\int_0^{\infty}\frac{F_{\rm exp}(x,\overline\gamma_{\rm E_0})\left[1-F_{\rm exp}^{M}(x,\overline\gamma_{\rm B_0})\right]}{1+x}dx \nonumber\\
    = &\frac{1}{\ln{2}} \sum\limits_{k=1}^{M}{M \choose k} (-1)^{k+1} \Delta{\mathcal{E}}\left(\A,\B\right),
    \label{eq:CsBTAS}
\end{align}
where $\Delta{\mathcal{E}}(\cdot)$ is an auxiliary function defined as
\begin{align}
    \Delta{\mathcal{E}} \left(A,B\right) = e^A {\rm E_{1}}(A) - e^B {\rm E_{1}}(B),
    \label{eq:deltaE1}
\end{align}
where ${\rm E_1}({\cdot})$ is the exponential integral function \cite[eq. (5.1.1)]{abramowitz1964handbook}.

Finally, we obtain the \ac{ASC} for the \ac{E-TAS} scheme by substituting \eqref{eq:etascdf} and \eqref{eq:eCDF} in \eqref{eq:Cs}, as: 
\newcommand\Ae{\frac{1}{\overline\gamma_{\rm B_0}}}
\newcommand\Be{{\frac{M}{\overline\gamma_{\rm E_0}} + \Ae}}
\begin{align}
    C_{\rm S}^{\rm E-TAS} = &\frac{1}{\ln 2}\int_0^{\infty}\frac{F_{\rm E}^{1}(x)\left[1-F_{\rm exp}(x,\overline\gamma_{\rm B_0})\right]}{1+x}dx \nonumber\\
    = &\frac{1}{\ln{2}} \Delta{\mathcal{E}}\left(\Ae,\Be\right),
    \label{eq:CsETAS}
\end{align}

Interestingly, expression \eqref{eq:CsETAS} has the exact same functional form as that originally deduced in Bloch's paper \cite[eq. 5]{Bloch2008} for the single-antenna case and Rayleigh fading. However, in the case of  \eqref{eq:CsETAS} we see that the eavesdropper's average \ac{SNR} is reduced by a factor of $M$. This implies that the effective \ac{SNR} of the eavesdropper is reduced as the number of antennas is increased, which is beneficial for physical layer security. 

\section{Numerical Results}\label{sec:results}

In this section we study the performance of the proposed schemes for physical layer security purposes. \ac{MC} simulations are included in all instances to double-check the validity of the analytical results.

\begin{figure}[t]
    \centering
    \includegraphics[width=1\columnwidth]{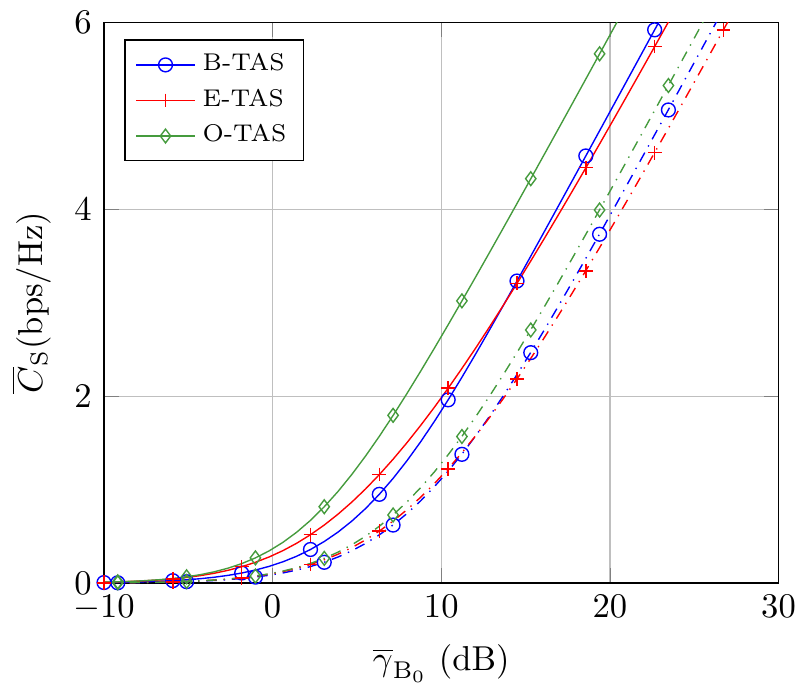}
    \caption{Average secrecy capacity $(\overline C_{\rm S})$ as a function of $\overline\gamma_{\rm B}$ for the schemes \ac{O-TAS}, \ac{B-TAS} and \ac{E-TAS} with $\overline\gamma_{\rm E_0}=10$ dB. The cases with $M=\{2,8\}$ antennas are represented with dash-dotted and solid lines, respectively. Markers correspond to \ac{MC} simulations.}
    \label{fig:Cs0}
\end{figure}

In Fig. \ref{fig:Cs0}, the \ac{ASC} is evaluated for all the three schemes, as a function of the reference average \ac{SNR} at the legitimate receiver $\overline\gamma_{\rm B_0}$ and a fixed value of $\overline\gamma_{\rm E_0}=10$ dB. The cases with $M=2$ and $M=8$ antennas are included with dash-dotted and solid lines, respectively. We see that the best possible performance is evidently obtained by the \ac{O-TAS} scheme, and that the performance gap with the sub-optimal schemes grows as $M$ is increased. One important observation is that the classical \ac{B-TAS} scheme only outperforms the newly proposed \ac{E-TAS} scheme for a sufficiently large $\overline\gamma_{\rm B_0}$. The performance gap between both sub-optimal schemes in the asymptotic regime is barely modified by increasing the number of antennas, and there is a range of \ac{SNR} values for which \ac{E-TAS} performs better than \ac{B-TAS}, as both the legitimate and eavesdropper's \acp{SNR} become comparable.

\begin{figure}[t]
    \centering
    \includegraphics[width=1\columnwidth]{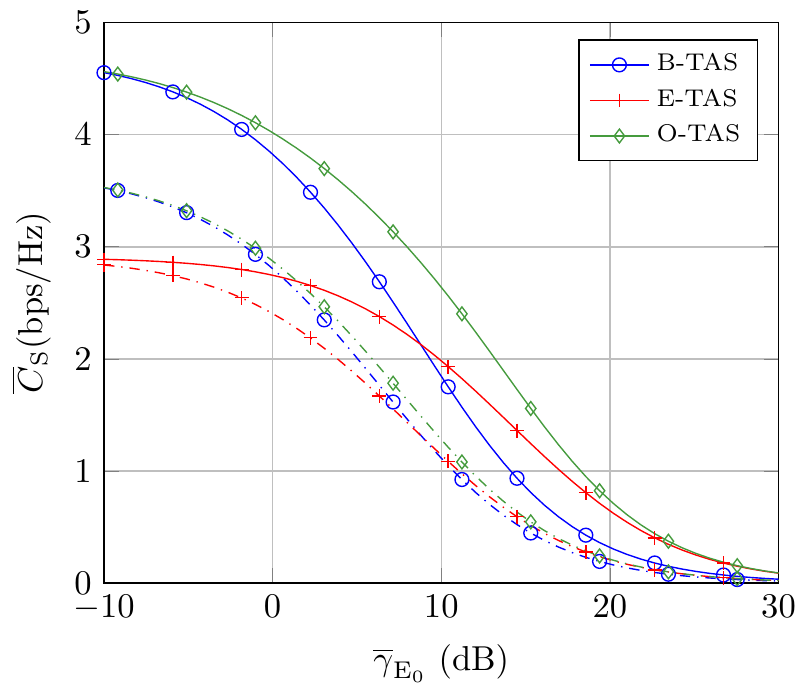}
    \caption{Average secrecy capacity $(\overline C_{\rm S})$ as a function of $\overline\gamma_{\rm E}$ for the schemes \ac{O-TAS}, \ac{B-TAS} and \ac{E-TAS} with $\overline\gamma_{\rm B_0}=10$ dB. The cases with $M=\{2,8\}$ antennas are represented with dash-dotted and solid lines, respectively. Markers correspond to \ac{MC} simulations.}
    \label{fig:Cs}
\end{figure}

In order to better investigate this behavior, we now evaluate in Fig. \ref{fig:Cs} the \ac{ASC} as a function of  $\overline\gamma_{\rm E_0}$, for a fixed value of $\overline\gamma_{\rm B_0}=10$ dB. Again, the cases with $M=2$ and $M=8$ are considered. We can see how the \ac{O-TAS} scheme acts as the upper-bound of the sub-optimal schemes. We observe that as the eavesdropper's \ac{SNR} grows, the \ac{E-TAS} scheme not only outperforms conventional \ac{B-TAS}, but also starts behaving closely to the \ac{O-TAS} scheme. At the extremes, where the \ac{SNR} value of one of the channels predominates, the sub-optimal scheme that better approximates \ac{O-TAS} is that whose predominant parameter is used in the selection criterion, see \eqref{eq:btas} and \eqref{eq:etas}.
Indeed, the \ac{ASC} is decreased in all instances as the eavesdropper's \ac{SNR} grows, and is increased when more antennas are used.

\begin{figure}[t]
    \centering
    \includegraphics[width=1\columnwidth]{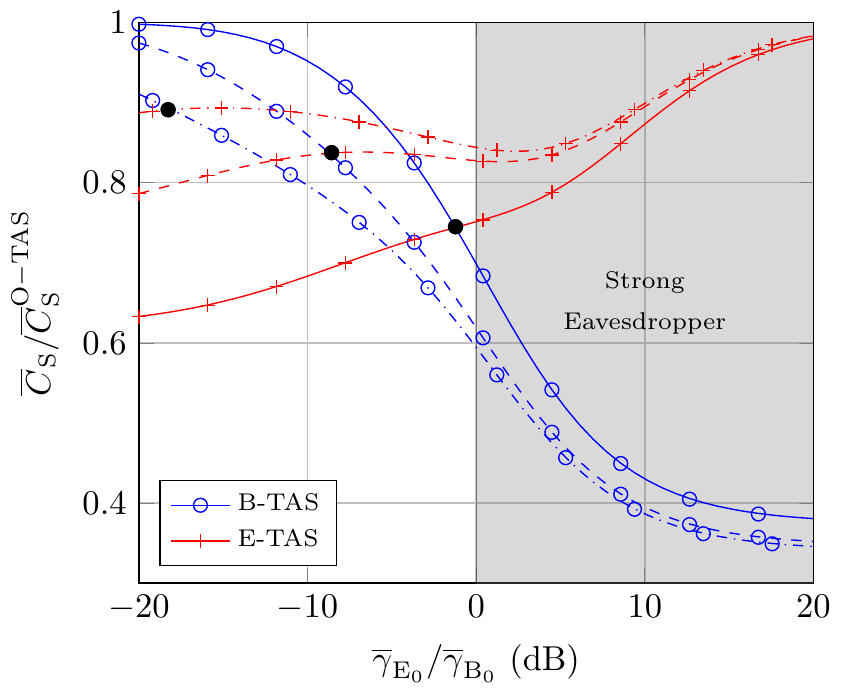}
    \caption{Normalized average secrecy capacity $(\overline C_{\rm S}/\overline C_{\rm S}^{\rm O-TAS})$ as a function of $\overline\gamma_{\rm E_0}/\overline\gamma_{\rm B_0}$, for different values of $\overline\gamma_{\rm B_0}=\{10,20,30\}$ dB. These correspond to the solid, dashed and dash-dotted lines in the figure, respectively, with $M=8$. Markers correspond to \ac{MC} simulations. The crossing SNR value is identified with a solid black marker for each pair of curves.}
    \label{fig:nCs}
\end{figure}

In order to evaluate the relative performance loss compared to the optimal \ac{TAS} scheme, and its interplay with the relation between the legitimate and eavesdropper's \acp{SNR}, we represent in Fig. \ref{fig:nCs}, the \acp{ASC} normalized to that of the \ac{O-TAS} case as a function of the ratio $\overline\gamma_{\rm E_0}/\overline\gamma_{\rm B_0}$. The case with $M=8$ antennas is used for exemplary purposes, and three different values for $\overline\gamma_{\rm B_0}$ are also used. As mentioned above, the best sub-optimal scheme varies considerably depending on the dominant channel, black dots are included to highlight when the best scheme changes from \ac{B-TAS} to \ac{E-TAS}. From the figure we can extract several insights: (\emph{i}) \ac{E-TAS} performs closer to the optimal scheme for a wider range of \ac{SNR} values; (\emph{ii}) \ac{E-TAS} is the best sub-optimal choice for the \emph{strong eavesdropper regime} \cite{vincentpoorstrongeve,zhang2020physical} (gray shaded area in the figure); (\emph{iii}) an adaptive transmission scheme able to switch between \ac{E-TAS} and \ac{B-TAS} criteria depending on the average \acp{SNR} and antenna-index information is recommended to enhance secrecy performance.


\section{Conclusion}\label{sec:conclusion}

We presented a new \ac{TAS} technique for \ac{PLS}, which is well-suited for scenarios on which antenna-quality information from the eavesdropper can be attained. By implementing a \ac{TAS} criterion which seeks to minimize the receive \ac{SNR} at the eavesdropper's side, we obtain noticeable performance gains for a wide range of \acp{SNR} compared to the conventional \ac{TAS} criterion classically used for \ac{PLS} when perfect \ac{CSI} is not available. Results also show that the performance of the proposed \ac{E-TAS} scheme without explicit \ac{CSI} knowledge is close to the optimal scheme with full \ac{CSI} in the strong eavesdropper's regime. The implementation of the \ac{E-TAS} technique may be well-suited for scenarios with strong eavesdroppers, and requires to have some sort of antenna quality indicator for the eavesdropper's channel. This can be done by exploiting the eavesdropper's RF leakages \cite{Mukherjee2012,Ghostbusters} and using energy detection techniques.




\section*{Acknowledgment}
The work of F.~Javier~L\'opez-Mart\'inez and Gonzalo J. Anaya-L\'opez was funded by Junta de Andaluc\'ia and the European Fund for Regional Development FEDER (project P18-RT-3175) and the Agencia Estatal de Investigaci\'on of Spain (PID2020-118139RB-I00).

\bibliographystyle{IEEEtran}
\balance
\bibliography{IEEEabrv,biblio}


\end{document}